\PassOptionsToPackage{pdfpagelabels=true}{hyperref} 

\documentclass[fleqn,usenatbib]{mnras}

\usepackage{newtxtext,newtxmath}
\usepackage[utf8]{inputenc}

\usepackage[T1]{fontenc}
\usepackage{ae,aecompl}

\usepackage{graphicx}	
\usepackage{amsmath}	
\usepackage{newtxtext,newtxmath}
\usepackage{multirow}
\usepackage[table,xcdraw]{xcolor}

\title[]{Assessing the spin-orbit obliquity of low-mass planets in the breaking the chain formation model: A story of misalignment} 

\author[L. Esteves et al.]{
Leandro Esteves,$^{1}$\thanks{E-mail: leandro.esteves@unesp.br}
André Izidoro,$^{2,3,4}$\thanks{E-mail: izidoro.costa@gmail.com}
Othon C. Winter,$^{1}$
Bertram Bitsch,$^{5}$ and
\newauthor
Andrea Isella$^{3}$
\\
$^{1}$UNESP, São Paulo State University, Grupo de Dinâmica Orbital e Planetologia, Guaratinguetá, CEP 12516-410, SP, Brazil\\
$^{2}$Department of Earth, Environmental and Planetary Sciences, 6100 Main MS 126,  Rice University, Houston, TX 77005, USA\\
$^{3}$Department of Physics and Astronomy, 6100 Main MS-550, Rice University, Houston, TX 77005, USA \\
$^{4}$Welch Postdoctoral Fellow \\ 
$^{5}$Max-Planck-Institut für Astronomie, Königstuhl 17, 69117 Heidelberg, Germany\\
}

\date{Accepted XXX. Received YYY; in original form ZZZ}

\pubyear{2022}

\begin{document}
\label{firstpage}
\pagerange{\pageref{firstpage}--\pageref{lastpage}}
\maketitle

\begin{abstract}
The spin-orbit obliquity of a planetary system constraints its formation history. A large obliquity may either indicate a primordial misalignment between the star and its gaseous disk or reflect  the effect of different mechanisms tilting  planetary systems after formation. Observations and statistical analysis suggest that system of planets with sizes between 1 and 4 R$_{\oplus}$ have a wide range of obliquities ($\sim0-30^{\circ}$), and that  single- and multi-planet transiting have statistically indistinguishable obliquity distributions. Here, we revisit the ``breaking the chains'' formation model with focus in understanding  the origin of spin-orbit obliquities. This model suggests that super-Earths and mini-Neptunes migrate close to their host stars via planet-disk gravitational interactions, forming chain of planets locked in mean-motion resonances. After gas-disk dispersal, about  90-99\% of these planetary systems experience dynamical instabilities, which spread the systems out.  Using synthetic transit observations, we show that if planets are born in disks where the disk angular momentum is virtually aligned with  the star's rotation spin, their final obliquity distributions peak at  about $\sim$5 degrees or less, and the obliquity distributions of single and multi-planet transiting systems are statistically distinct. By treating the star-disk alignment as a free-parameter, we show that the obliquity distributions of single and multi-planet transiting systems only become statistically indistinguishable if planets are assumed to form in primordially misaligned natal disks with a  ``tilt'' distribution peaking at $\gtrsim$10-20 deg. We discuss the origin of these misalignments in the context of star formation and potential implications of this scenario for formation models.
\end{abstract}




\section{Introduction}

The angle between a star's spin axis and its planets' orbital total angular momentum -- typically refereed as spin-orbit obliquity or  ``stellar obliquity'' -- is a window into the system formation and dynamical evolution history. The Sun's obliquity  is about 6 degrees~\citep{beckgiles05,souamsouchay12}, suggesting that the solar system planets formed in a gaseous disk rotating in a broadly common sense with the Sun\footnote{For a discussion on the origin of the solar system obliquity see, for instance, \cite{baileyetal16}, \cite{gomesetal17}, \cite{lai16} and references there in.}. 

Although the Sun's obliquity has been precisely determined  from the proper motion of sunspots, helioseismology data, and  high-precision ephemerides of solar system objects~\citep{beckgiles05,souamsouchay12}, measuring the obliquity of distant planet-host stars is relatively way more challenging. This is  because the angular resolution of ordinary observations may not be able to resolve details on the spatial scale of the stellar surface~\citep{albrechtetal22}. Yet, stellar obliquities of exoplanetary systems have been successfully estimated for more than $\sim$100 stars via different techniques \citep[see][]{triaud18,albrechtetal22} as the $\nu \sin{i_{\star}}$ method~\citep{schlaufman10,walkowiczetal13,mortonwinn14,winnetal17}, the photometric variability method~\citep{Mazehetal15},  the Rossiter–McLaughlin method~\citep{rossiter1924,McLaughlin1924,quelozetal00,gaudietal07,kunovacetal21}, the asteroseismic method~\citep{chaplinetal13,huberetal13,VanEylenetal14,campanteetal16},  the spot-crossing anomalies method~\citep{desertetal11,sanchis-Ojedaetal11,mazehetal15b,feietal18}, the gravity-darkening method~\citep{barnes09,masuda15}, and by combining astrometry with orbital solutions obtained from stellar radial-velocity curves~\citep{sahlmannetal11,sahlmannetal11b}.

The  obliquities of the best-characterised planet host-stars, show a wide range of values~\citep[e.g.][]{hebrardetal08,winnetal09,quelozetal10}, extending from almost perfectly-aligned systems~\citep[e.g. Kepler-30;][]{sanchis-ojedaetal12} to systems with very misaligned planets~\citep[e.g. Kepler-63 and Kepler-56 with obliquities of $\sim$87.8 and $\sim$45 degress, respectively; ][]{sanchis-ojedaetal13,huberetal13}.  If misaligned planets are systematically born from  primordially misaligned gaseous  disks or, alternatively, from well-aligned disks but get tilted later~\citep[e.g.][]{wumurray03,rogers12,cebronetal13,petrovich15,andersonlai18} remains unclear. The goal of this paper is to compare the obliquity distribution of observed systems with that produced from planet formation models accounting for planetary growth, gas-driven planet migration, and dynamical instabilities~\citep{bitschetal19,izidoroetal21,izidoroetal22,bitschizidoro23}. Before getting into  the details of our model and approach, we briefly discuss some of the latest developments in the field in terms of spin-orbit obliquity measurements in order to motivate our study (see a for detailed review see \cite{albrechtetal22}).

\subsection{The obliquity distribution  of exoplanetary systems}

To the end of measuring exoplanetary obliquities, as most  methods and techniques rely on transit-observations, our initial knowledge on the obliquity of exoplanetary systems has mainly revolved around giant planet systems -- hot Jupiter systems --  simply because they are  easier to detect than relatively smaller planets~\citep[e.g.][]{win10}. The since ever-growing exoplanet population and the  observational evidence that planets with sizes between those of the Earth and Neptune are far more common than gas giant giants~\citep{mayoretal11,howardetal12,fressinetal13,petiguraetal13,zhuetal18,muldersetal18} has more recently driven a series of studies focused, instead, on characterising the obliquity distribution of low-mass planetary systems~\citep{mortonwinn14,winnetal17,munhozperets18,loudenetal21}. These studies have mostly taken advantage of the  large number of stars with radii and rotational periods available in the Kepler catalogue and the fact that the $\nu \sin{i_{\star}}$ method is relatively low-cost. It requires less spectral resolution and sensitivity  than the Rossiter–McLaughlin method~\cite[e.g.][]{munhozperets18} and does not require observations to be taken at the timing of transit~\citep{gaudietal07}. Most of the analysis and subsequent discussion presented in this work will make use of results coming from the $\nu \sin{i_{\star}}$ method and will focus on stars hosting low-mass planets which represent the bulk of the Kepler data.

The obliquity of planetary systems constrain planet formation models.  The mean obliquity of a selected sample of 156 Kepler stars with reliably measured photometric periods has been constrained from the $\nu \sin{i_{\star}}$ method to be smaller than 20 degrees~\citep{winnetal17}, with 99\% confidence level. This result is in reasonable agreement  with the results of ~\cite{munhozperets18} who found - using 257 California Kepler Survey star targets~\citep{johnsonetal17,petiguraetal17} --   a mean obliquity for the entire sample of 20 degrees, with a spreading of 10 degrees. Although a few percent of the stars  in the samples of ~\citet{winnetal17} and ~\cite{munhozperets18}  also host giant planets, \cite{loudenetal21} have performed a dedicated analysis for systems of low-mass planets ($R<4 R_{\oplus}$) only. By selecting  153 host stars, these authors found mean obliquities varying between 37 and 58 degrees. By splitting their stellar sample into groups of ``hot'' ($>$6250~K) and ``cold'' ($<$6250~K) stars, these authors found that the mean obliquities of cold stars are lower, varying between 18 to 38 degrees, whereas hot stars have higher mean obliquities, varying from 48 to 88 degrees.

Another interesting result come from the  obliquity distribution of single and multi-planet transiting systems. Statistical analysis of 70 Kepler host stars by \citet{mortonwinn14} initially suggested that the  obliquities of systems showing single and multiple transiting planets are statistically distinct. In a follow-up study, with increased sample size, \cite{winnetal17} showed that this trend has vanished and the authors  concluded that the best-fit obliquity distributions do not depend on the transit multiplicity. This finding is also supported by the analysis of \cite{munhozperets18}. 

Overall, statistical studies suggest that the obliquities of exoplanetary systems (with no hot-Jupiters) may be reasonably modest, with derived upper limits of up to 30 degrees~\citep{winnetal17,munhozperets18}. The best-fit distributions of ~\cite{munhozperets18} and ~\cite{loudenetal21} --  which suggest mean obliquities of about 20 degrees -- are particularly interesting. A mean value of $\langle\psi\rangle\sim$20 degrees is about 3 times higher than the solar system obliquity, which may suggest that our planetary system has a fairly atypical obliquity compared to those in the California Kepler Survey sample~\citep{johnsonetal17,petiguraetal17}. It remains  elusive whether \textit{the breaking the chain model} is consistent with obliquities derived from observations analysis, or if it requires additional ingredients or physics to be reconciled with observational constraints.

The goal of this paper is to assess the obliquity distribution of planets produced in the breaking the chain migration model~\citep{izidoroetal17,izidoroetal21,estevesetal22,izidoroetal22,bitschizidoro23}. Our planet formation model is designed to produced low-mass (or small) planets, i.e, planets with masses lower than 20~$M_{\oplus}$ (or  $R <4R_{\oplus}$)  and with orbital periods shorter than $\sim$100 days. These planets are usually referred to as hot super-Earths and/or mini-Neptunes. We will also include in our analysis simulations that produce close-in small planets along with cold outer gas giant planets~\citep{bitschizidoro23}.
This is motivated by the fact that about $\sim$40\% of systems of close-in small planets host also gas giant planets at large orbital distances ($M \sim M_{{\rm Jup}}$ and $P>365$~days; ~\cite{barbatoetal18,zhuetal18,bryanetal19,rosenthaletal22}). In this work, we will address two main questions: i) What is the obliquity distribution of exoplanets produced in \textit{the breaking the chain model} and how does it compare to observations? ii) Does \textit{the breaking the chain model} predict a statistically indistinguishable  obliquity distribution for single and multi-planet transiting systems as suggested by recent studies?

The layout of the paper is as follows. In \S \ref{sec:sim} we describe our planet formation model and the geometry of the obliquity problem.  In \S \ref{sec:obliquityproblem} we present our main results. In \S \ref{sec:giantplanets} we show the influence of including systems hosting distant giant planets on our main results. Finally, in \S \ref{sec:summary} we summarise our results and discuss the major implications of our findings.

\section{Simulations}\label{sec:sim}

\textit{The breaking the chain model} suggest that super-Earths and mini-Neptunes grew quickly and migrated during the gas disk phase, via planet-disk gravitational interactions \citep{cresswellnelson06,terquempapaloizou07,paardekooperetal11}, forming resonant chains of planets anchored at the disk inner edge~\citep{massetetal06,romanovalovelace06,romanovaetal21}. Shortly after gas disk dispersal, from $\sim$90 to 99\% of systems become dynamically unstable leading to a phase of orbital crossing and giant impacts~\citep{izidoroetal17,izidoroetal21,izidoroetal22}. This scenario is broadly consistent with multiple observational constraints, as the period ratio distribution of Kepler planets, the peas-in-a-pod feature~\citep{weissetal18}, the exoplanet radius valley~\citep{fultonetal17}, and the observed planet multiplicity distribution. 

In this work, we do not perform new simulations, instead  we revisit two sets of simulations of \textit{the breaking the chain model} which have been shown to be consistent with these observational constraints~\citep{izidoroetal21,izidoroetal22,bitschizidoro23}. Each of our different set of  simulations mix a fraction of  ``stable'' and ``unstable'' systems. Following previous studies~\cite[e.g.][]{izidoroetal17}, stable systems are defined as those where planets remained in resonant chains after gas disk dispersal. Unstable systems are those where  planets experienced dynamical instabilities after gas disk dispersal. Our two set of simulations mix 2\% stable systems and 98\% unstable systems. 

Our nominal set of simulations is designed to produce exclusively low-mass planets -- planets with masses lower than 20$M_{\oplus}$ or radii smaller than 4$R_{\oplus}$. For a detailed description of the initial conditions and model setup, we refer the reader to the model-III of  \cite{izidoroetal21}.  Our second set of simulations consists of a mix of simulations producing  two types of planetary systems. This second set of simulations comes from \cite{bitschizidoro23}, and mix 60\% of systems with only low-mass planets and 40\% systems of low-mass planets with cold gas giant planets also known as cold-Jupiters (CJs). The chosen mixing fraction is motivated by observational analysis, suggesting that about 40\% of the close-in low mass planets also host gas giants~\cite[e.g.][]{rosenthaletal22}. 

In the next section, we describe the geometry of the spin-orbit obliquity  problem and our approach to calculate system obliquities using these simulations.

\subsection{The obliquity problem}\label{sec:obliquityproblem}

In the simulations of \cite{izidoroetal21} and \cite{bitschizidoro23}, the central star is modelled as a point-mass object, and there is no priori or required assumption on the star spin orientation. In this work, in order to calculate obliquities we have to assume star-spin orientations for their systems. In order to avoid a costly approach by re-running their simulations, we treat the star's spin orientation as a free parameter of the model. We take each planetary system produced in  simulations of \cite{izidoroetal21} and \cite{bitschizidoro23} and assign a star-spin orientation (sampling from a given distribution) for each system via post-processing of the data. 

We represent the star spin orientation in a reference system conveniently defined. The x-y plane of our reference system coincides with the underlying gas disk mid-plane of the simulations of \cite{izidoroetal21} and \cite{bitschizidoro23}. Consequently, the z-direction of our reference system is also parallel to the angular momentum vector of the gaseous disk (Figure \ref{fig:obliquity_scheme}a). For each of our planetary systems, the star's spin orientation is represented by the unit vector $\hat{n}_{\textrm{*}}$, which is uniquely defined by the angles $\theta$ and $\Lambda$, showed in Figure \ref{fig:obliquity_scheme}a. We recall that 
the orientation of $\hat{n}_{\textrm{*}}$ is a free parameter in our model and  with the definition of this angle, we can now proceed and calculate the spin-orbit obliquity of our systems.

\noindent{
\begin{figure}
	\includegraphics[scale=1]{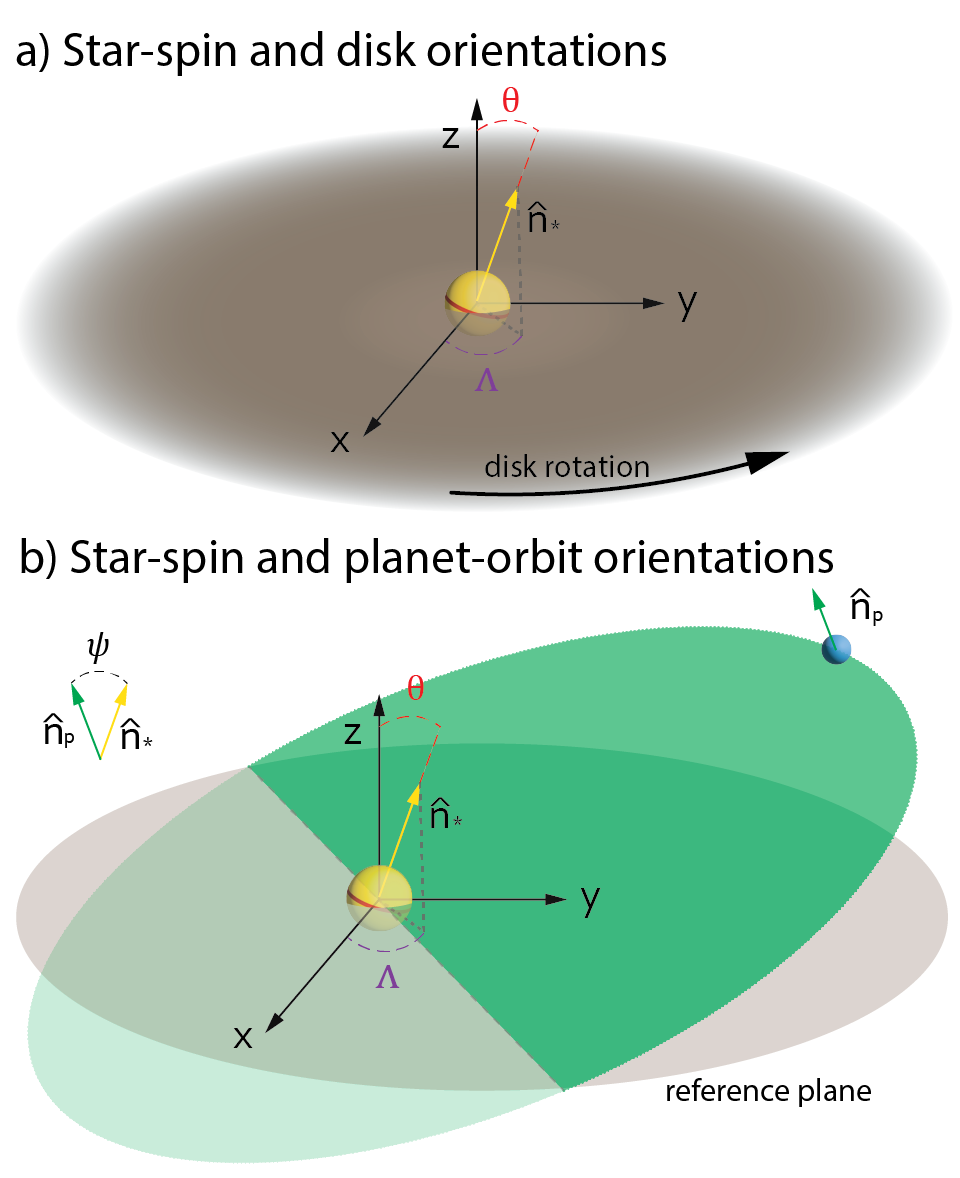}
    \caption{\textbf{a)} Schematic of star-disk orientations. The star is shown as a yellow sphere at the center of the coordinate system. $\hat{n}_{\textrm{*}}$ is the norm  of the star spin vector. $\theta$ is  the angle  between $\hat{n}_{\textrm{*}}$ and the z-axis. $\Lambda$ is the azimuthal angle defined by the projection of $\hat{n}_{\textrm{*}}$ onto the x-y plane and the x-axis. The protoplanetary disk angular momentum is parallel to the z-axis. The gas disk mid-plane is shown in dark-grey.
    \textbf{b)} Star-planet orientations. The orbital plane of the planet is shown in green, which is this case, for illustration purposes do not coincides with the original disk mid-planet. The planet's angular momentum is represented by the vector $\hat{n}_{\textrm{p}}$. The spin-orbit obliquity is given by $\psi$, which is defined as the angle between $\hat{n}_{\textrm{*}}$ and $\hat{n}_{\textrm{p}}$, see inset on top left of (b). We show a single planet for simplicity. In the case, of multiple planets, the obliquity is defined using the total orbital angular momentum of all transiting planets.}
    \label{fig:obliquity_scheme}
\end{figure}
}

The spin-orbit obliquity -- thereafter refereed to as $\psi$ -- is defined as the angle between the stellar spin axis and the orbital angular momentum of the planet (or total orbital angular momentum of planets). Note that the directions of the angular momentum vector of the planet (or planetary system) and that of the gaseous disk are not necessarily the same. Figure \ref{fig:obliquity_scheme}b shows
a geometric representation of $\psi$ for a star hosting a single planet, for simplicity. The grey-plane in Figure \ref{fig:obliquity_scheme}b represents the  disk-midplane shown in Figure \ref{fig:obliquity_scheme}a. $\hat{n}_{\textrm{p}}$ is the unit vector representing the orbital angular momentum of the plane, which is an output of our planet formation simulations. The calculation of the system obliquity becomes relatively easy in our method because we have information on the orbital configuration of the planet (or planets in a real case). This is one of the major advantages of our approach compared to that used by transit observations, which relies on the geometry of the transit to estimate obliquities. In Figure \ref{fig:obliquity_scheme}, $\hat{n}_{\textrm{*}}$ is the unit vector representing the star's spin axis (defined via $\theta$ and $\Lambda$). Finally, the obliquity angle ($\psi$) can be calculated as
\begin{equation}\label{eq:obl_def}
    \cos \psi = \hat{n}_{\textrm{p}}\cdot \hat{n}_{\textrm{*}}.
\end{equation}

In our subsequent analysis, we assume that the star's spin orientation may vary from perfectly-aligned to very misaligned configurations relative to the orientation of the natal gaseous disk angular momentum. We will assume different distributions and ranges of  $\theta$ and $\Lambda$. Next, we compare the distributions of obliquities produced by these different distributions of $\theta$ and $\Lambda$ with the obliquity distribution inferred from observations, with particular focus on the distributions of single and multi-planet transiting systems.

The motivation for this approach is two-fold. First, disk observations indeed suggest that stars may be primordially misaligned with their young disks \citep[e.g.][]{krausetal20,Bietal20,kuffmeieretal21}. Second, numerical simulations also suggest that primordial disk-star misalignment can be a natural outcome of chaotic accretion during star formation~\citep{bateetal10,fieldingetal115,bate18}, disk's magnetic warping~\citep{foucartlai11,laietal11,romanovaetal21}, disk tilt induced by inclined external companions like stars~\citep[e.g.][]{lubow00,batygin12,batyginadams13,lai14,spaldingbatygin14,bate18,cuelloetal19,cuelloetal20} or planets~\citep[e.g.][]{matsakoskonigl17} and torques emerging from the disk itself~\citep{epsteinetal22}. 

However, the distribution of primordial disk-star orientations is not strongly constrained by observations and neither by theoretical studies. From a theoretical point of view, the degree of primordial (mis-)alignment may come at many different levels.  For instance, \citet{bate18} and \citet{fieldingetal115} used hydrodynamical simulations to study the evolution of protostars and circumstellar discs in a star cluster. Their findings suggest that the star-disc misalignments exceed 30 degrees for about 50\% of the protostars, reaching misalignments of up to $\sim90$ degrees. \citet{romanovaetal21} used magnetohydrodynamical simulations to show that gas accretion onto stars with strong magnetic fields  may produced misalignment ranging from 5 degrees up to 40 degrees. We will later show that our simulations seems to favour mean values of $\theta$ larger than about 10-20 degrees.

Note that in our post-processing analysis -- by imposing star-spin orientations to fully formed planetary systems --  we neglect any  effects of the primordial misalignment on the migration and growth of planets in the gaseous disk and after gas disk dispersal.  This is probably an important caveat of our work, in particular, if the star is oblate~\cite[e.g.][]{spalding19}. Accounting for these effects in a self-consistent way would require a complete new set of simulations. This study motivates future studies addressing this issue. However, at the same time, we believe that our simplified approach can be used to draw  conclusions that will be also broadly valid in this more sophisticated scenario. These issues will be addressed in a follow-up work.

\subsection{Tilting stars}

The  obliquity distribution of exoplanets is traditionally either represented by Fisher distributions~\citep{fisher53,fisher93}, as proposed by \cite{fabryckywinn09,munhozperets18}, or by Rayleigh distributions~\citep{winnetal17}. As discussed before,  the obliquity of a planetary system from our simulations can be directly calculated once we assume an orientation for the star spin, by setting values for $\Lambda$ and $\theta$. In order to have a statistical sample, we adopted Rayleigh distributions to represent $\theta$, without loss of generality. We have verified that our main results are also qualitatively consistent with exponential and Fisher distributions. We assume that $\Lambda$ follows an uniform distribution, as would be expected from a random orientation of stars in space~\citep[e.g.][]{winnetal17}, but later in the paper, we explore the response of our results to the choice of $\Lambda$.

\noindent{
\begin{figure}
	\includegraphics[scale=0.58]{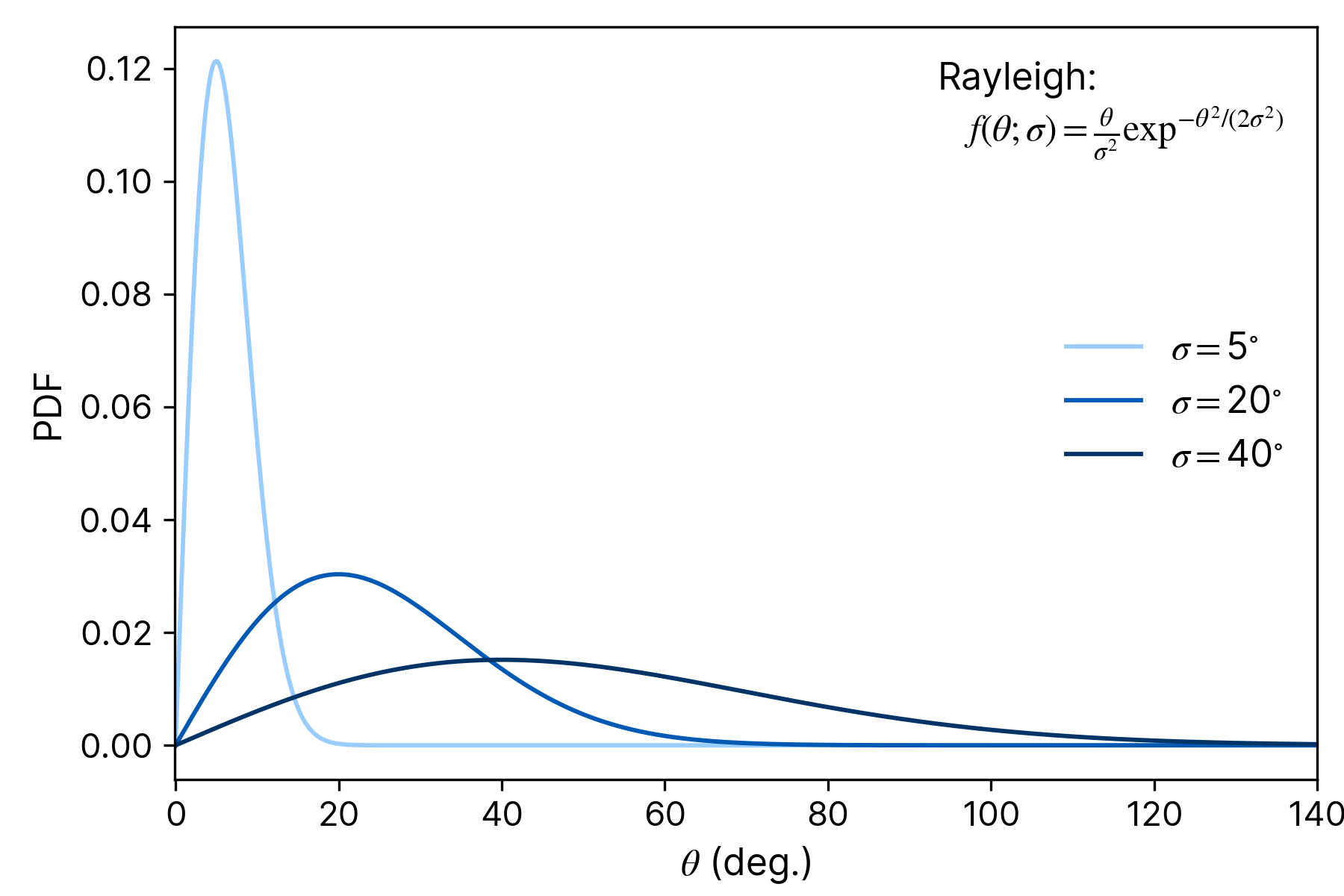}
    \caption{Probability density functions of $\theta$ following Rayleigh distributions with different scale parameters $\sigma$, as indicated by the colours. Light-blue shows $\sigma=5$, blue shows $\sigma=20$, and dark-blue $\sigma=40$ degrees, respectively.}
    \label{fig:prob_dist}
\end{figure}
}

Figure \ref{fig:prob_dist} shows examples of  probability density functions (PDF) of Rayleigh distributions of $\theta$ for different scale parameters $\sigma$ (the mean star tilt relative to its natal disk; see $\theta$ in Figure \ref{fig:obliquity_scheme}), set to 5, 20 and 40 degrees. For each value of scale parameter,  we produce a distribution of $\hat{n}_{\textrm{*}}$, that is assigned to our simulated planetary systems and  used to calculate their respective obliquity distribution.

\subsection{Simulated transit observations}

In order to effectively  compare the results of our model with those estimated from  observations, we have to add observational biases to our systems by  performing synthetic transit observations of them. It is important to include biases in our sample because the transit probability decreases as the planet semi-major axis increases. In addition, systems with multiple planets may have planets missed in transit observations due to mutual orbital inclinations, which may impact the overall planet multiplicity distribution, and consequently the obliquity distribution~\citep{izidoroetal17,izidoroetal21}. 

We follow the approach of \citet{izidoroetal21} when conducting synthetic transit observations. Each simulated planetary system is observed from many different lines of sight evenly spaced by 0.1 degree from angles spanning
from 90 to -90 degrees relative to the x-y plane of Figure \ref{fig:obliquity_scheme}.
Azimuthal viewing angles are evenly spaced by 1 degree, and are assumed to vary from 0 to 360 degrees.  For each line of sight where at least one planet transits, we store the  physical and orbital parameters of  transiting planets  creating a synthetic observed system. A single simulated planetary system when observed from multiple lines of sight may create many synthetic observed systems. We aim at comparing the obliquity distributions of synthetic observed systems with that inferred from real Kepler observations. 

To calculate the obliquity of systems produced from our synthetic observations, we compute the total orbital angular momentum of the system $\hat{n}_{\textrm{p}}$ accounting only for observed the planets for a given line of sight. We randomly assigned to each synthetically observed system a star-spin orientation ($\hat{n}_{\textrm{*}}$) by drawing $\theta$ from a Rayleigh distribution with mode $\sigma$. In our nominal analysis, we assume that $\Lambda$ follows an uniform distribution between 0 and 360 degrees, but we also test the effects of restricting the size of this interval in Section \ref{sec:results}.

\section{Results}\label{sec:results}

\subsection{Comparing the obliquity distribution of simulations and observations}\label{sec:comparision}

\noindent{
\begin{figure*}
	\includegraphics[scale=0.85]{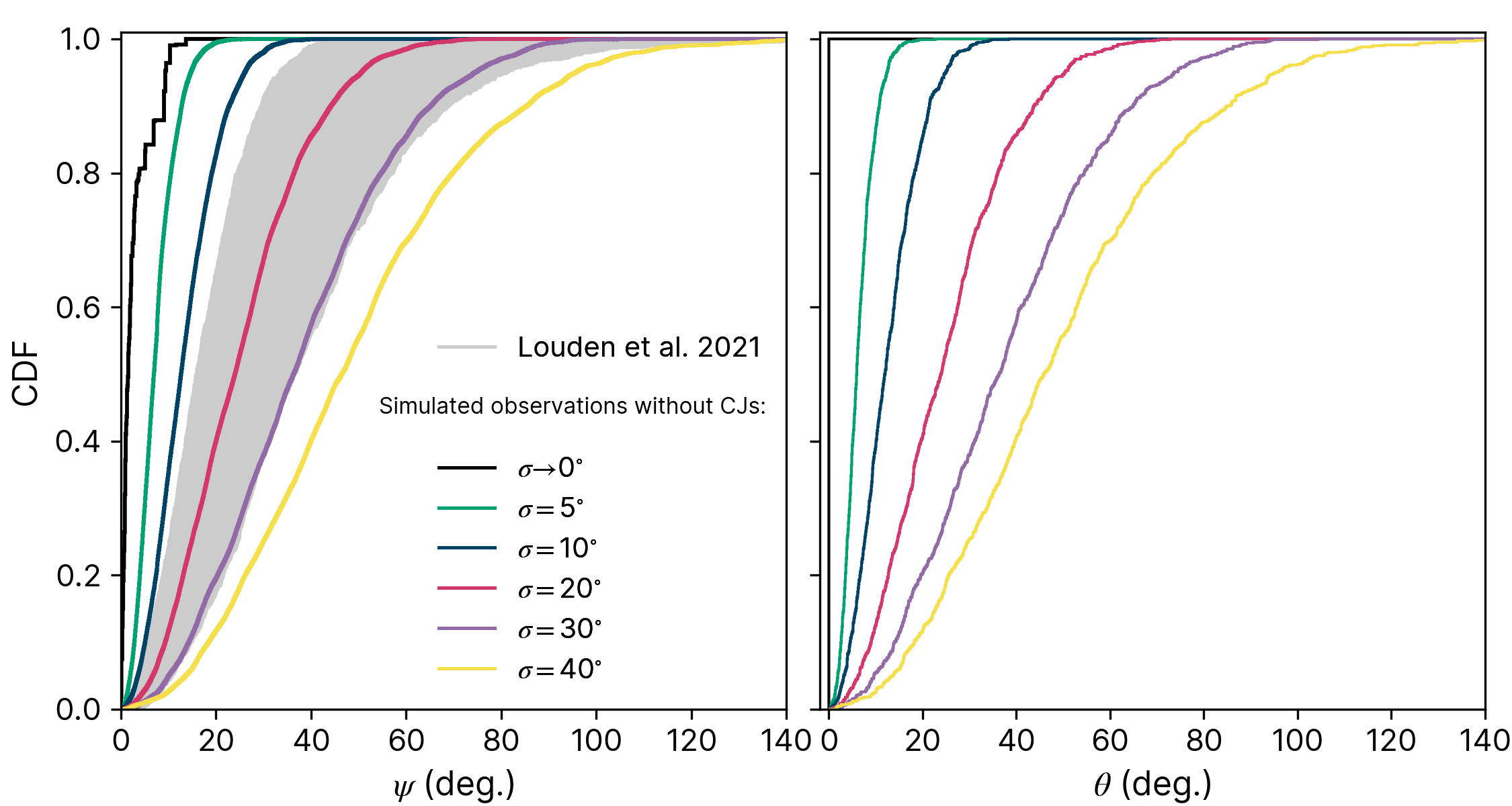}
    \caption{\textbf{Left panel} Cumulative distributions of obliquity ($\psi$)  of  simulated planetary systems (with no cold gas giants) and  real observations. Grey lines  show the cumulative distributions of Kepler exoplanets as inferred by \citet{loudenetal21}. In black and in colourful lines, we show the obliquity distributions of our simulated planetary systems for different values of $\sigma$. {\bf Right panel:} Cumulative distribution functions of $\theta$ for different values of $\sigma$. }
    \label{fig:compare}
\end{figure*}
}

Figure \ref{fig:compare} (right panel) shows the  obliquity distributions of our synthetically observed systems when we assume different star-spin orientation distributions ($\theta$) via the scale parameter $\sigma$. This figure includes only systems where all planets have masses lower than $\sim$20$M_{\oplus}$, and no cold gas giants exists. We recall that for each value of $\sigma$ we generate a different Rayleigh distribution of $\theta$, which are shown on the left panel of Figure \ref{fig:compare}. The case where $\sigma \to 0^{\circ}$, represented by the black solid line, corresponds to the scenario where the star spin is perfectly aligned with its natal disk angular momentum direction, and all stars in the sample have $\theta=0^{\circ}$. In the left panel of Figure \ref{fig:compare}, we plot in gray the estimate obliquity distribution of real observations. We take as a proxy for observations the results of  \citet{loudenetal21}, which were derived for a sample of Kepler systems with planets smaller than 4~$R_{\oplus}$ and without hot Jupiters.

As discussed earlier, \citet{loudenetal21} suggest that hot ($>6250$~K) and cold ($<6250$~K) stars  have statistically different obliquities distributions. As in all our numerical simulations  the central star is a solar-mass star, we decided to compare our results with the distribution of obliquities of stars with effective temperature lower than 6250~K, instead of using their entire sample. An effective temperature of $<6250$~K is consistent with G-type stars, as our Sun. The mean obliquities of stars in the cold sample of \citet{loudenetal21} varies between  $\left \langle \psi \right \rangle = 18^{\circ}$ and $38^{\circ}$.  We will take this range as reference to conduct our subsequent analysis.  The gray lines of the left panel of Figure \ref{fig:compare} show 2500 Rayleigh distributions plot altogether (forming a filled area), where the scale parameters corresponds to values varying between $18^{\circ}$ and $38^{\circ}$.

Figure \ref{fig:compare} shows that when the host star is assumed perfectly aligned with its natal disk ($\sigma\rightarrow0$, $\theta=0$), the obliquities of our planetary systems are systematically lower than 10 degrees, with typical values of only a few degrees (see black solid line of Figure \ref{fig:compare}). In this scenario, the obliquities of our planetary systems are much lower than the mean obliquity estimated by the best-fit distributions of observation analysis (grey lines). This trend remains  valid for all scenarios where $\sigma$  is lower than 10 degrees. This strongly suggests that  dynamical instabilities after gas disk dispersal alone cannot account for the estimated obliquity distribution of low-mass exoplanets, as currently estimated by best-fit distributions. Our results better match  these estimate when stars are assumed primordially misaligned with their natal disks, with a typical tilt ($\theta$) of about $\sim$20 degrees or more. We now analyse how this picture changes when our systems include also cold gas giant planets.

\subsection{The influence of external gas giant planets}\label{sec:giantplanets}

In this section, we present the results of our set of simulations where 40\% of the planetary systems host also cold gas giant planets at distances larger than 1~au~\citep{bitschizidoro23}. This set of simulations is particularly interesting for our analysis because of two reasons. Firstly, because dynamical instabilities after the gas disk dispersal tend  to be relatively more violent in systems with cold giant planets, sculpting the architecture of the inner system and reducing planet multiplicity~\citep{bitschizidoro23}. Secondly, because secular perturbations from distant gas giants may have a strong impact on the orbital inclination of the inner system~\citep{hansen17,mustilletal17,beckeradams17,laipu17,pulai18,bitschetal19,bitschetal20,rodetlai21}, and therefore, the inner system obliquity.

\noindent{
\begin{figure}
	\includegraphics[scale=0.7]{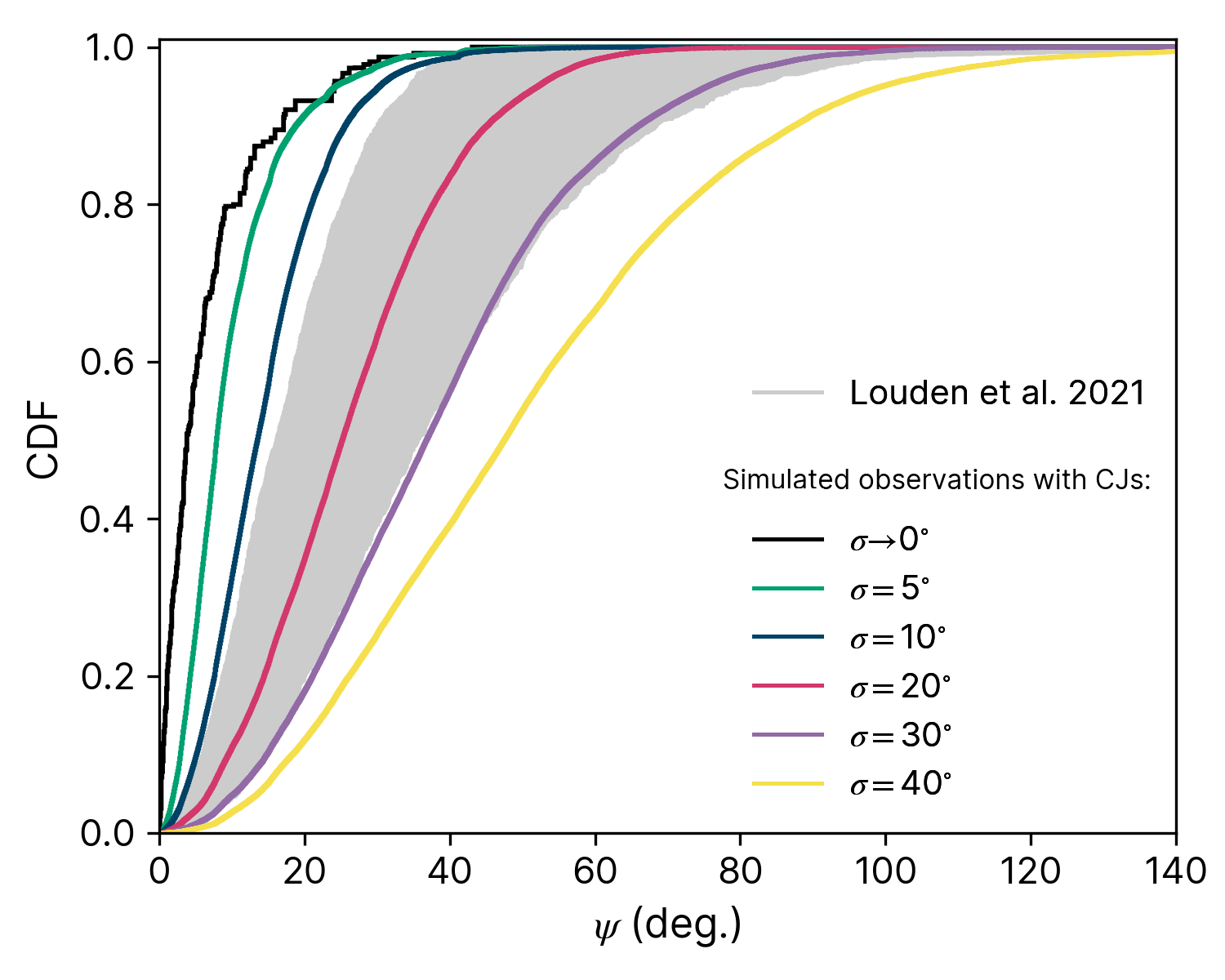}
    \caption{Same as left panel of Figure \ref{fig:compare}, but for simulated observations of planetary systems containing cold gas giants.}
    \label{fig:compare_giant}
\end{figure}
}

Figure \ref{fig:compare_giant} shows the obliquity distribution produced from  simulations with cold gas giants.  In black, we show the distribution when the star spin is assumed perfectly aligned with the  disk total angular momentum ($\sigma\rightarrow0$, $\theta=0$). As it may be expected, the obliquity distribution of systems with cold gas giants  is relatively wider than those where systems contain only low-mass planets. The solid black line shows systems with obliquities of up to 40 degrees, compared to a maximum value of about $\sim$10 degrees when only low-mass planets are present in the systems. However, it is still clear that the obliquity  of these systems is relatively much lower than that suggested by observational analysis of ~\cite{loudenetal21} when $\sigma$ is lower than 20 degrees. Therefore, we conclude that the presence of cold gas giant planets alone is not enough to increase the obliquity of planetary systems to levels estimated by the best-fit distribution derived from observation analysis. As in the case without cold gas giants, the best match to the obliquity distribution derived from  observations analysis require planets to be born from primordially misaligned disks, with mean tilts of about $\sim$20 degrees or more.

\subsection{The obliquity distribution of single and multi-planet transiting systems}

\noindent{
\begin{figure}
	\includegraphics[scale=0.7]{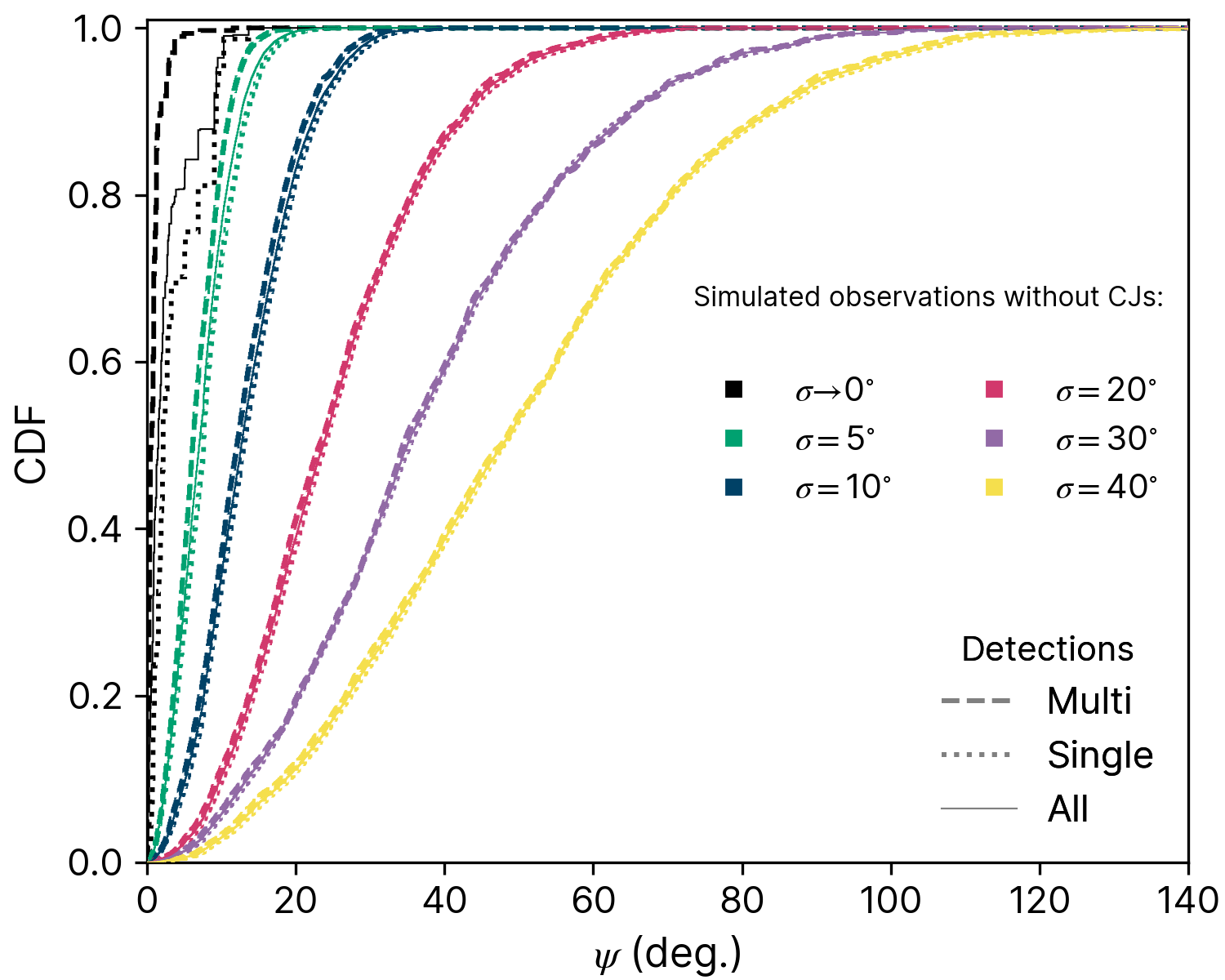}
    \caption{Cumulative distributions of obliquity ($\psi$) of simulated planetary systems divided in single and multi-planet transiting systems. Dashed lines show distributions of multi-planet transiting systems. Dotted lines represent single-transit systems. Solid lines contains all systems. Lines are colour-coded as indicated by the parameter $\sigma$. These simulations do not contain cold gas giants.}
    \label{fig:obl_dist}
\end{figure}
}

\noindent{
\begin{figure}
	\includegraphics[scale=0.7]{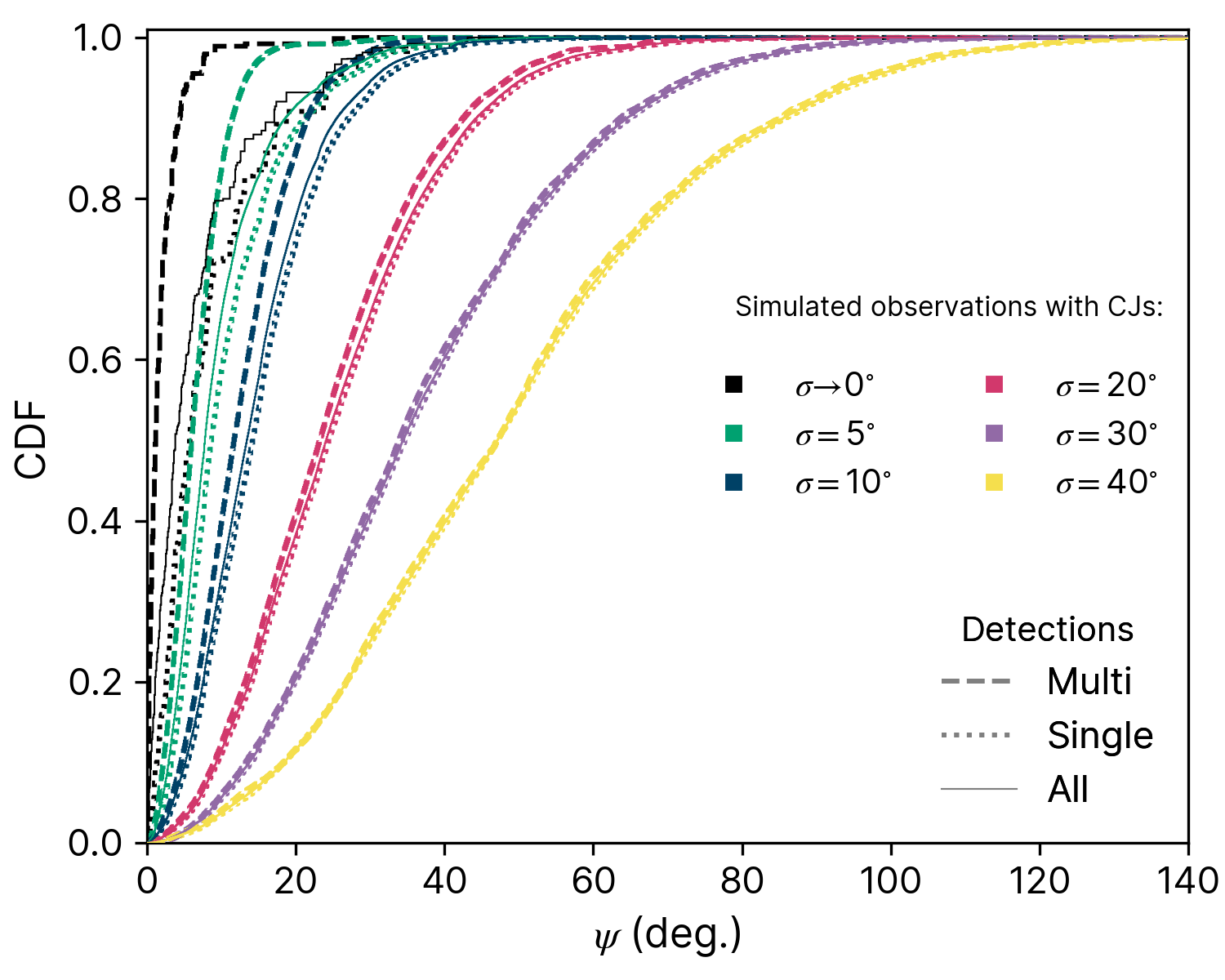}
    \caption{Same as of Figure \ref{fig:obl_dist}, but for simulations containing cold gas giants.}
    \label{fig:obl_giant}
\end{figure}
}

We now turn our attention to the obliquity distributions of single and multi-planet transiting systems. Our goal here is to compare the results of our model and those of observations analysis suggesting that single and multi-planet transiting systems have  statistically indistinct distributions of obliquity~\citep{winnetal17}.

Figure \ref{fig:obl_dist} shows the cumulative obliquity distributions divided in groups of single and multi-planet transiting systems ($N_{\rm obs}\geq2$). This figure corresponds to simulations that produce only low-mass planets. It shows that when $\sigma\to 0$, the obliquities of multi-planet transiting systems (black dashed line of Figure \ref{fig:obl_dist})  are systematically lower than that of single transiting ones (black dotted line). This is because single transiting planets tend to  come mainly from system of multiple planets that have relatively higher mutual orbital inclinations. This increase of probability of multiple planets being missed during transit simulations and of only one planet being detected by transit. Previous studies have indeed suggested that a significant fraction of Kepler single transiting planets are in fact not truly single, but a simple outcome of transit observations missing planets with mutual orbital inclinations~\citep{izidoroetal17,muldersetal18,izidoroetal21}.

Figure \ref{fig:obl_giant}, which shows the results of our set of simulations  with cold gas giants, shares a similar trend to that of Figure \ref{fig:obl_dist}. An important difference, however, is that the case with cold gas giants, single and multi-planet transiting systems have notably more different distributions of obliquity. Overall, Figures \ref{fig:obl_dist} and \ref{fig:obl_giant}  show that, as we increase $\sigma$, the obliquity distributions of single and multi-planet transiting systems tend towards  more similar distributions until they eventually become visually identical. In order to rigorously test whether and under which values of $\sigma$ these distributions become statistically indistinct, we use the Kolmogorov-Smirnoff statistical test (KS). As the results of KS tests are  sensible to sample sizes, we also explore the robustness of result as a function of the sample size.  It is particularly important to keep in mind that the sample sizes of observational analysis contains about  $\sim100$ stars~\citep{winnetal17}. In our analysis, we randomly select sub-samples of our entire sample of synthetic observations. We have performed our tests considering sub-sample sizes of 50, 100, 1000 and 10000 systems. For each sub-sample size, we repeat the KS-test 100 times by re-drawing from the full sample and we report the mean p-values calculated over all 100 trials.

Table \ref{tab:p-values} and \ref{tab:p-values2} show that, regardless of the sample size, when $\sigma\to 0$ our two set of simulations produce obliquity distributions of single and multi-planet transiting systems statistically distinct. This is not surprising, and confirm our previous arguments and visual analysis. Sample sizes of 50 and 100 systems, with $\sigma$ equal to 5 or 10 degrees, accept the null hypothesis that the obliquities of single and multi-planet transiting systems are drawn from the same distribution. However, as we increase the sample size,  the power of the test increases suggesting that, in reality, only systems with $\sigma\gtrsim20$ degrees produce statistically indistinct obliquity distributions for single and multi-planet transit systems.

\noindent{
\begin{figure}
	\includegraphics[scale=0.7]{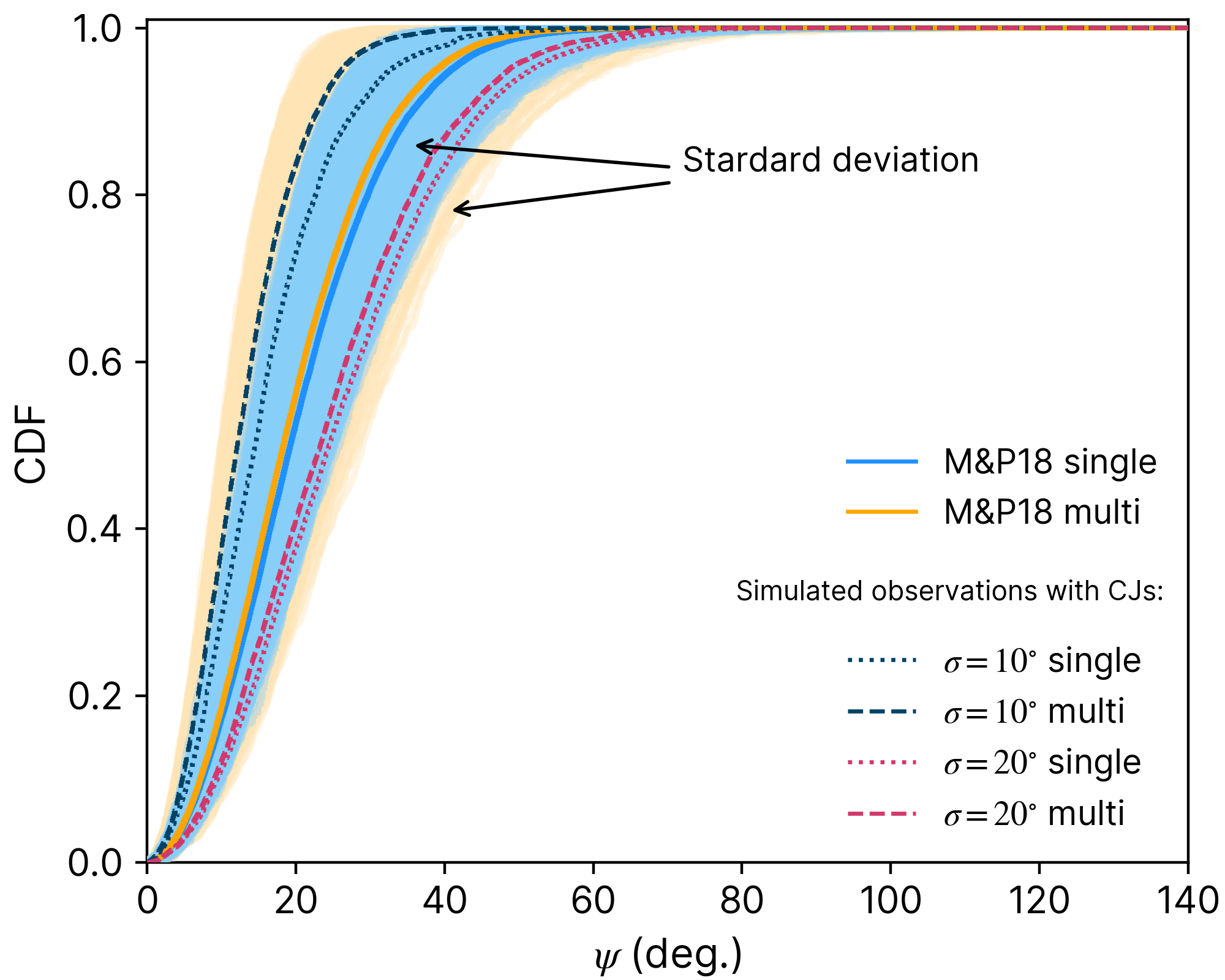}
    \caption{Cumulative distributions of obliquity ($\psi$) for single and multi-planet transiting systems comparing our simulated systems with data inferred from real observations. Blue and orange lines show the mean obliquity of single and multi-planet transiting systems as inferred by \citet{munhozperets18}. The light-blue and light-orange distributions are the standard deviation of single and multi, respectively. The dark-blue and red lines denotes our simulated systems with best-matching $\sigma$, where the dotted lines illustrates single systems and dashed lines multi systems.}
    \label{fig:compare_MnP18}
\end{figure}
}

This result is particularly interesting because it corroborates with the best-fit results of observational analysis~\citep{munhozperets18,loudenetal21} suggesting the obliquity distribution of low-mass exoplanets  peak at about 20 degrees. Our results suggest that tilts following a Rayleigh distribution with $\sigma = 20$ degrees or so,
between the star spin and its natal disk implies that singles and multi-planet transiting systems should indeed have statistically indistinguishable  distributions of obliquity.

It is important to keep in mind that we use as reference for this analysis and discussion the best-fit solutions of previous works~\citep{winnetal17,munhozperets18}. We have used  KS-tests to confirm that the best-fit obliquity distributions of single and multi-planet transiting systems derived by \citet{munhozperets18} are statistically indistinguishable, regardless of sample size (p-values are larger than 0.3). However, best-fit solutions estimated from real observations come with  large associated uncertainties. The standard-deviations  of the best-fits solutions of \citet{munhozperets18} are shown in Figure \ref{fig:compare_MnP18} (see blue and yellow regions). Large uncertainties  may affect our ability to distinguish single and multi-planet transiting systems. This may be particularly important if the obliquity distribution of observations (singles and multis together) is eventually constrained to be in fact consistent with very low values of $\sigma$ (e.g. $ \sigma = 5$ deg or lower; see Figure \ref{fig:obl_giant}). In this hypothetical scenario -- which is not favoured by current observational analysis ~\citep{winnetal17,munhozperets18,loudenetal21} -- our results predict that single and multi-planet transiting systems should be statistically distinct, yet observation analysis may not be able to distinguish them. On the other hand, if $\sigma$ is larger than  $\sim$20 degrees, our results suggest that we can confidently conclude that single and multi have indistinguishable obliquity distributions.

\begin{table*}
\centering
\begin{tabular}{lclccclc}
\hline
\multicolumn{1}{c}{}                         &                               &                      & \multicolumn{5}{c}{Rayleigh ($\sigma$)}                                                                                                                                                      \\
\multicolumn{1}{c}{\multirow{-2}{*}{Sample}} & \multirow{-2}{*}{$\sigma \to 0$} &                      & 5                             & 10                            & 20                            & \multicolumn{1}{c}{30}                            & 40                            \\ \hline
50                                           & \cellcolor[HTML]{FD6864}0.000     &                      & \cellcolor[HTML]{67FD9A}0.308 & \cellcolor[HTML]{67FD9A}0.531 & \cellcolor[HTML]{67FD9A}0.573 & \cellcolor[HTML]{67FD9A}0.598                     & \cellcolor[HTML]{67FD9A}0.519 \\
100                                          & \cellcolor[HTML]{FD6864}0.000     &                      & \cellcolor[HTML]{67FD9A}0.197 & \cellcolor[HTML]{67FD9A}0.467 & \cellcolor[HTML]{67FD9A}0.545 & \cellcolor[HTML]{67FD9A}0.465                     & \cellcolor[HTML]{67FD9A}0.559 \\
1000                                         & \cellcolor[HTML]{FD6864}0.000    &                      & \cellcolor[HTML]{FD6864}0.000     & \cellcolor[HTML]{67FD9A}0.150 & \cellcolor[HTML]{67FD9A}0.406 & \cellcolor[HTML]{67FD9A}0.465                     & \cellcolor[HTML]{67FD9A}0.442 \\
10000                                        & \cellcolor[HTML]{FD6864}0.000     & \multicolumn{1}{c}{} & \cellcolor[HTML]{FD6864}0.000     & \cellcolor[HTML]{FD6864}0.000     & \cellcolor[HTML]{67FD9A}0.429 & \multicolumn{1}{c}{\cellcolor[HTML]{67FD9A}0.434} & \cellcolor[HTML]{67FD9A}0.425 \\ \hline
\end{tabular}
\caption{\label{tab:p-values} Mean p-values of KS-tests comparing the obliquity distributions of single and multi-planet transiting systems for simulations without cold gas giant planets. The columns show the sample size and the different values of $\sigma$, from 0 to 40 degrees. The sample is  randomly selected from our full set of simulated observations and the KS-test is repeated 100 times when computing each mean p-value. The background colors of table cells  are shown in red for p-values lower than 0.05, and in green for p-values larger than 0.05. For p-values higher than 0.05 we confirm the null hypothesis that the obliquities of single and multi-planet transiting systems are drawn from the same distribution.}
\end{table*}

\begin{table*}
\centering
\begin{tabular}{lclccclc}
\hline
\multicolumn{1}{c}{}                         &                               &                      & \multicolumn{5}{c}{Rayleigh ($\sigma$)}                                                                                                                                                      \\
\multicolumn{1}{c}{\multirow{-2}{*}{Sample}} & \multirow{-2}{*}{$\sigma \to 0$} &                      & 5                             & 10                            & 20                            & \multicolumn{1}{c}{30}                            & 40                            \\ \hline
50                                           & \cellcolor[HTML]{FD6864}0.000    &                      & \cellcolor[HTML]{67FD9A}0.272 & \cellcolor[HTML]{67FD9A}0.505 & \cellcolor[HTML]{67FD9A}0.577 & \cellcolor[HTML]{67FD9A}0.541                     & \cellcolor[HTML]{67FD9A}0.561 \\
100                                          & \cellcolor[HTML]{FD6864}0.000     &                      & \cellcolor[HTML]{67FD9A}0.086 & \cellcolor[HTML]{67FD9A}0.454 & \cellcolor[HTML]{67FD9A}0.548 & \cellcolor[HTML]{67FD9A}0.535                     & \cellcolor[HTML]{67FD9A}0.577 \\
1000                                         & \cellcolor[HTML]{FD6864}0.000     &                      & \cellcolor[HTML]{FD6864}0.000     & \cellcolor[HTML]{FD6864}0.049 & \cellcolor[HTML]{67FD9A}0.461 & \cellcolor[HTML]{67FD9A}0.461                     & \cellcolor[HTML]{67FD9A}0.476 \\
10000                                        & \cellcolor[HTML]{FD6864}0.000     & \multicolumn{1}{c}{} & \cellcolor[HTML]{FD6864}0.000     & \cellcolor[HTML]{FD6864}0.000     & \cellcolor[HTML]{67FD9A}0.053 & \multicolumn{1}{c}{\cellcolor[HTML]{67FD9A}0.301} & \cellcolor[HTML]{67FD9A}0.466 \\ \hline
\end{tabular}
\caption{\label{tab:p-values2} Same as Table \ref{tab:p-values}, but for simulated observations containing cold gas giants.}
\end{table*}

\subsection{``Singles'' and  ``Multis'' obliquity distributions: How/Why do they become  statistically indistinguishable?}

In order to understand why the distributions of single and multi-planet transiting systems become statistically indistinguishable for large values of $\sigma$ we have explored the response of the obliquity distributions of planetary systems to two geometric aspects of our model. We study the response of $\psi$ to: i) different levels of orbital inclination of  systems of multi and singles; and ii) the assumed distribution of $\Lambda$ (see Figure \ref{fig:obliquity_scheme}).  We start our analysis with scenario i).

Although when studying the scenario i) the ideal approach would consist of performing new  simulations designed to produce planetary systems with different levels of orbital inclination, we anticipate to the reader that this type of numerical simulation is computationally expensive. Therefore, in order to make this study doable in a reasonable time-frame, we decided to create  toy ``planetary systems'' that qualitatively mimic the outcome of real simulations. 

When creating our new toy systems, we do not assign inclinations to individual planets themselves, but instead we create ``systems'' with different levels of inclinations by setting the system invariant plane inclination (plane perpendicular to the planets' total orbital angular momentum). We create three set of systems, which we refer to as the ``low'', ``medium'', and ``high'' orbital inclination cases. We assume that within each of these categories systems come in two flavours, as single and multi-planet transiting systems. Single planet systems in a given category have relatively higher orbital inclinations than their counterparts. This choice is naturally motivated by the results of our real simulations (see Figures \ref{fig:obl_dist} and \ref{fig:obl_giant}).

The inclination distributions of the ``low'' inclination case are generated following  exponential distributions with scale parameters set to  $\beta_{\textrm{s}} = 4.45^{\circ}$ and $\beta_{\textrm{m}} = 1.04^{\circ}$ for single- and multi-planet transiting systems, respectively (the subscript ``s'' stands for single and  ``m'' for multi-planet transiting systems) . The distributions of the ``medium'' inclination case correspond to $\beta_{\textrm{s}} = 8^{\circ}$ and $\beta_{\textrm{m}} = 2^{\circ}$. Finally, the distribution of the ``high'' inclination case has $\beta_{\textrm{s}} = 16^{\circ}$ and $\beta_{\textrm{m}} = 4^{\circ}$. By setting  the inclination of invariant planet of our toy systems, we can calculate their obliquity distributions using the same procedure used for our nominal simulations.

\noindent{
\begin{figure*}
	\includegraphics[scale=0.8]{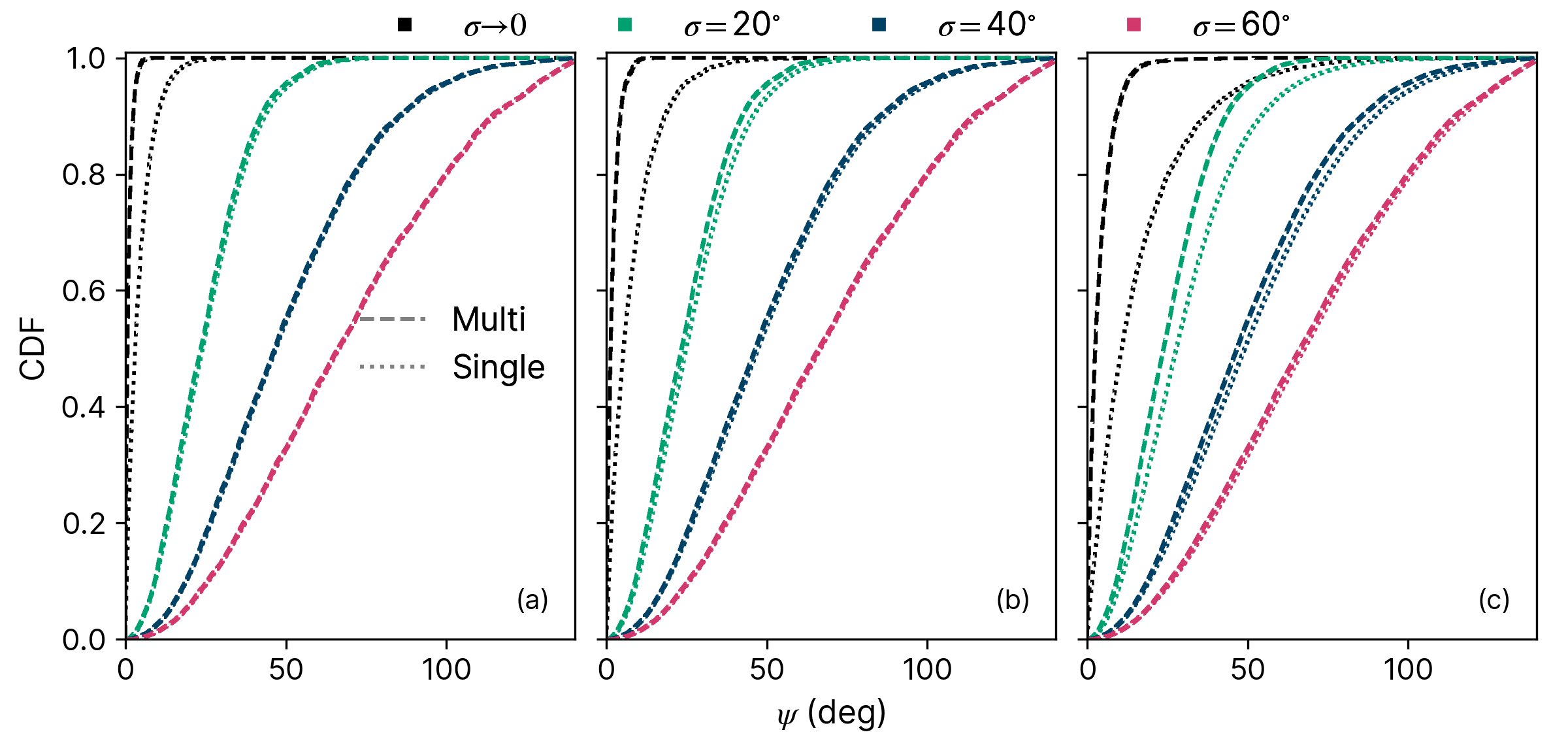}
    \caption{Response of the obliquity distributions of single and multi-planet transiting systems to the inclination of the observed planetary systems. Cumulative distributions are generated from our toy-models planetary systems. Panel a) shows our  ``low'' inclination case, b) shows the ``medium'' inclination case, and c) the ``high'' inclination case. The black lines shows the obliquity distributions for  $\sigma \to 0$. Colour-coded lines shows the distributions for different values of $\sigma$.}
    \label{fig:toy_model1}
\end{figure*}
}

Figure \ref{fig:toy_model1} shows the obliquity distributions of the ``low'' (Figure \ref{fig:toy_model1}a),  ``medium'' (Figure \ref{fig:toy_model1}b), and ``high'' (Figure \ref{fig:toy_model1}c) inclination cases. In all three panels, the distributions of singles and multi-planet transiting system are shown as dotted and dashed lines, respectively. It is clear that for $\sigma\rightarrow0$, the distribution of singles and multis are distinct in all panels, with the ``high'' case having, itself, relatively more distinct distributions of singles and multis. The colour-coded lines of Figure \ref{fig:toy_model1}  shows that, as in the case of our nominal simulations, the distributions of single and multis converge towards more similar distributions as we increase $\sigma$. Yet, there is another clear trend: the higher the level of inclination of the toy systems is, the larger the value of $\sigma$ needed  to make the distributions of singles and multis converge towards a indistinguishable distribution. 

\noindent{
\begin{figure*}
	\includegraphics[scale=0.68]{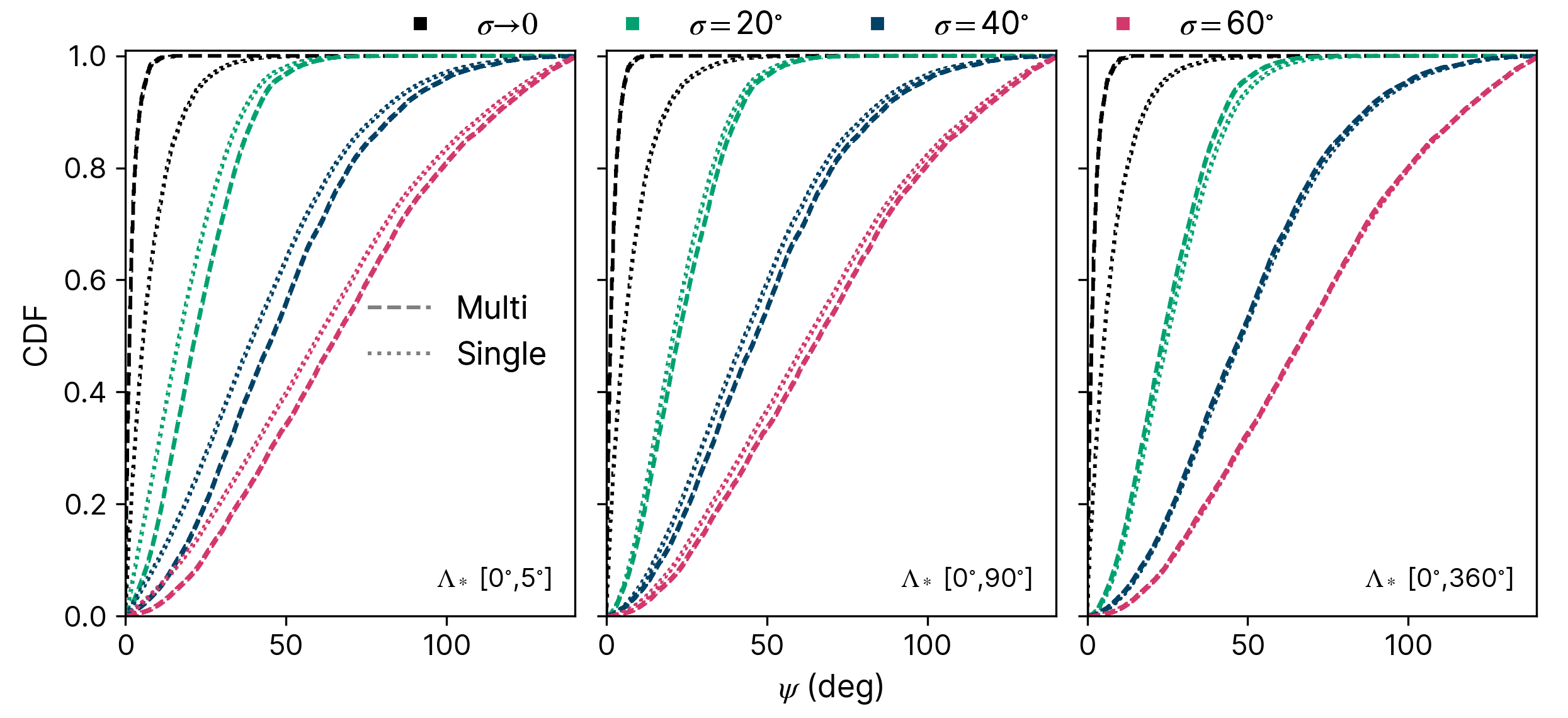}
    \caption{Response of the obliquity distributions of single and multi-planet transiting systems to the assumed distribution of $\Lambda$. Cumulative distributions are generated from our toy-model planetary systems. In our nominal analysis $\Lambda$ follows an uniform distribution varying between 0 and 360 degrees.  From left-to-right, the panels shows the distributions of obliquities when we sample $\Lambda$ following uniform distributions within the following intervals [0°, 5°],  [0°, 90°] and  [0°, 360°].}
    \label{fig:toy_model2}
\end{figure*}
}

We now analyse the response of the obliquity distributions of single and multi-planet transiting systems to the assumed distribution of $\Lambda$. We recall that in our nominal analysis $\Lambda$ is a free-parameter of the model and is drew from an uniform distribution sampled between 0 and 360 degrees. We now limit the range of $\Lambda$ to understand its effect on the obliquity distributions of single and multi-planet systems. Figure \ref{fig:toy_model2} shows that  the  obliquity distributions of singles and multi-planet transiting systems only become statistically indistinguishable if $\Lambda$ is drawn from a broad distribution of values.  If $\Lambda$ is limited, for instance, to values ranging between 0 and 90 degrees (or less), not even larger values of $\sigma$ yield distributions of single and multi-planet transiting systems statistically indistinct (Figure \ref{fig:toy_model2} left-panel). This suggests that star-spin orientations should not only be tilted by a mean value of $\gtrapprox$20 degrees, relative to their natal disk orientation, but the ``tilt'' should also follow an isotropic distribution in the azimuthal direction (e.g. Figure \ref{fig:toy_model2} right panel). This is a required condition to ensure that single and multi-planet transiting systems have indistinguishable obliquity distributions in our model. It implies that there is no correlation between the orientation of the plane of the gaseous disk and the azimuthal orientation of the stellar spin.

\section{Summary}\label{sec:summary}

In this work, we revisit simulations of the \textit{the breaking the chain model} to study the formation of systems of hot super-Earths and mini-Neptunes \citep{izidoroetal21,bitschizidoro23} with focus on the obliquity distribution of these planetary systems. 

The so-called breaking the chain model suggests that close-in super-Earths and mini-Neptunes migrated via tidal-interaction with their natal disk forming long chains of planets anchored at the disk inner edge, with planets locked in first order mean motion resonances. After gas disk dispersal, about 90-99\% of the resonant system become dynamically unstable leading to a phase of orbital crossing and giant impacts. This instability phase shapes the dynamical architecture of planetary systems. This scenario has been demonstrated to be  broadly consistent
with a number of constraints of Kepler observations. This includes the broad period ratio distribution of adjacent planet pairs, planet multiplicity distribution,
the exoplanet radius valley~\citep{fultonetal17}, and the peas-in-a-pod feature~\citep{weissetal18}. Simulations of \cite{bitschizidoro23} includes also the effects of cold gas giant planets ($P>365$ days), which  observational analysis suggest to exist in about 40\% of the systems of super-Earths and mini-Neptunes. In this work, we expand on previous studies by focusing on the study of the spin-orbit obliquity distribution of systems produced in the \textit{the breaking the chain model} and how it compares to that estimated from observations. 

Statistical analysis of Kepler data suggests that the mean obliquity of systems of hot super-Earths and mini-Neptunes is reasonably high, with best-fit distributions suggesting mean values of about $\sim$20 degrees or so \citep{winnetal17,munhozperets18,loudenetal21}. It has been also proposed that the obliquity distribution of single and multi-planet transiting are statistically indistinguishable~\citep{winnetal17,munhozperets18}. In this paper we tested if \textit{the breaking the chain formation model} is also consistent with the  results of these analysis.

We show that if systems of super-Earths and mini-Neptunes are born from gaseous disks well aligned with the star spin orientation (e.g. disk angular momentum and and star's spin parallel to each other), then their mean obliquity distribution  should peak at about 5 degrees or less.  Our best match to the best-fit obliquity distribution of exoplanets~\citep{winnetal17,munhozperets18,loudenetal21} come from planet formation scenarios where we assume that gaseous disks are primordially misaligned with their host stars. Our results favour a primordial star-disk misalignment distribution that peaks at about $\gtrsim$10-20 degrees. This scenario is consistent with statistical analysis suggesting that single and multi-planet transiting systems have broadly indistinguishable obliquity distributions~\citep{winnetal17,munhozperets18}. Our results suggest that most planetary systems may be born with a level of misalignment up to a few times higher than that of our current solar system, that is about 7 degrees. The origin of these misalignments remains unclear. It is probably either associated to the process of star formation and accretion~\citep{bateetal10,laietal11,fieldingetal115,bate18,romanovaetal21}, or to the effects of external perturbers tilting the disk or entire planetary systems. The later may include the effects of stars during the stellar cluster phase, star companions, passing stars during the system long-term evolution~\citep{lubow00,batygin12,batyginadams13,lai14,spaldingbatygin14,bate18,cuelloetal19}, or the effects of highly inclined and eccentric giant planets changing disk orientation~\citep[e.g.][]{bitschetal13}.

In  our entire analysis, we neglect potential gravitational effects on the star, gaseous disk, and planets arising from  our envisioned primordial star-disk misalignment scenarios. Yet, we speculate that primordial misalignments may have a key role triggering dynamical instabilities of resonant chains after gas disk dispersal. We would naively expect that the rate of instability to correlate with the degree of primordial misalignment, in particular, if stars are slightly oblate during the early stages of planet formation~\citep{spaldingbatygin16,spaldingetal18}.  Stellar quadrupole potential  may have an important role destabilising planetary systems~\citep{spaldingbatygin16}. This would lead to the prediction that pristine systems of resonant chains should preferentially come in systems with low spin-orbit obliquities. Long resonant systems departing from this rule, would indicate that the host stars were virtually spherical during their early stages.

Finally, our planet formation simulations suggest that secular perturbations of cold gas giants on systems of inner super-Earths and mini-Neptunes {\it alone} cannot account for the actual estimated obliquity distribution of low-mass exoplanets. It  requires planets to be born in primordially misaligned disks, planetary systems to be tilted by stars during the embedded stellar cluster phase, or via other mechanisms.

\section*{Acknowledgements}
The authors wish to thank the anonymous referee for their valuable comments and feedback about the manuscript.
L.~E. acknowledges the financial support from FAPESP through grant 2021/00628-6. A.~Izidoro acknowledges support via NASA grant 80NSSC18K0828 (to Rajdeep Dasgupta). A.~Izidoro and A.~Isella are grateful to the The Welch Foundation for support via grant No. C-2035-20200401.
O.~C.~W. thanks Conselho Nacional de Desenvolvimento Científico e Tecnológico (CNPq) proc. 305210/2018-1 and FAPESP proc. 2016/24561-0 for financial support.
O.~C.~W and A.~I thank the Brazilian Federal Agency for Support and Evaluation of Graduate Education (CAPES), in the scope of the Program CAPES-PrInt, process number 88887.310463/2018-00, International Cooperation Project number 3266.
B.~B., thanks the European Research Council (ERC Starting Grant 757448-PAMDORA) for their financial support.

\section*{Data availability}
The data underlying this article will be shared on reasonable request to the corresponding author.




\bibliographystyle{mnras}
\bibliography{references} 





\bsp	
\label{lastpage}
\end{document}